
\input phyzzx
\nonstopmode
\nopubblock
\sequentialequations
\twelvepoint
\overfullrule=0pt
\tolerance=5000
\input epsf

\line{\hfill IASSNS-HEP 95/71}
\titlepage
\title{Statistical Transmutation and Phases of
Two-Dimensional Quantum Matter}\foot{Invited talk at the
150th anniversary of Boltzmann's birth, Academi Lincei, Rome,
June 1994.}

\author{Frank Wilczek\foot{Research supported in part by DOE grant
DE-FG02-90ER40542.~~~wilczek@sns.ias.edu}}
\vskip.2cm
\centerline{{\it School of Natural Sciences}}
\centerline{{\it Institute for Advanced Study}}
\centerline{{\it Olden Lane}}
\centerline{{\it Princeton, N.J. 08540}}
\endpage

\abstract{After surveying the quantum kinematics and
dynamics of statistical transmutation, I show how this concept
suggests a phase diagram for the two-dimensional matter in a
magnetic field, as a function of quantum statistics.  I discuss
the fundamental properties of quasiparticles in the different
phases, and briefly suggest {\it gedanken\/} -- but not manifestly
infeasible --
experiments to show up these properties.}

\endpage

\REF\disclaim{This manuscript for the most part sticks close
to the talk as given.  It is meant to be a light but sophisticated
introduction for non-experts.  Accordingly, and because I am writing
it in isolated conditions, I have not attempted to supply a scholarly
set of references.  Please do not assume that any specific unfamiliar
remark found here is original, though some are.
Section 4 on quasiparticle propagation is based on
recent collaborative work done with my students R. Levien and
C. Nayak, available as IASSNS-HEP 94/108 and 95/67.}

I feel my subject may be something of an anomaly here, since
it does not grow in any very direct way out of Boltzmann's great
work on statistical physics, and
indeed involves at its core concepts that
were totally unknown in Boltzmann's day.
However in a broader sense it
is an appropriate homage, since it speaks, in a way that is both
direct and I think quite beautiful,
to Boltzmann's overarching ambition
to relate microscopic laws to macroscopic behavior.  So without
further ado, let me describe for you how fundamental new possibilities
for microscopic laws governing identical particles in quantum mechanics
underly some spectacular recent discoveries in condensed matter physics
[\disclaim ].

\chapter{Rules of Quantum Statistics}

\section{Quantum Kinematics: Exchange and Braiding}


The fundamental specification of the
dynamics of a quantum-mechanical system must be a rule
for calculating transitions amplitudes.  Let the amplitude
for a process wherein $n$ particles at initial
positions
$(r^{(I)}_1, \dots ,r^{(I)}_n)$ arrive at positions
$(r^{(F)}_1, \dots ,r^{(F)}_n)$ through path $\lambda$
be denoted
$$
A(r^{(I)}_1, \dots ,r^{(I)}_n; r^{(F)}_1, \dots ,r^{(F)}_n ; \lambda )
$$
or $A( I\rightarrow F; \lambda)$ or even $A(\lambda )$ if no ambiguity
arises.  The usual way to specify transition amplitudes is in terms
of an action functional of the form
$$
A(\lambda ) ~=~
  \exp \bigl(
 i\int^{(F)}_{(I)} \Pi dr_j  ~
     {\cal L} (r_j (\tau ), \dot{r}_j (\tau )) \bigr)~.
\eqn\amp
$$
The complete amplitude is obtained by summing over all paths,
which gives the Feynman path integral.  This general form has the
advantage that it automatically incorporates the rule for
composing amplitudes for consecutive processes, that is
$$
A(\lambda_{1 +2}) ~=~ A(\lambda_1 ) A(\lambda_2 )~,
\eqn\composing
$$
where the path $\lambda_{1+2}$ is obtained by
first following path
$\lambda_1$ and then path $\lambda_2$.
If, in addition, ${\cal L}$ is taken to be the classical Lagrangian,
then the system obeys the correspondence principle.

The correspondence principle guides us in finding the laws
of quantum dynamics, but it
is not determinative, since it certainly can happen that different
quantum systems have the same classical limit.

New possibilities arise, in particular, when the space of paths is
topologically non-trivial -- that is, when there are different sorts
of path that cannot be continuously deformed one into another.
(Here I am speaking, of course, of paths in $n$-particle position
space.  When I wish to speak of the trajectories of individual
particles, I will call them world-lines.)  The
classical equation of motion
can be reproduced by comparing only infinitesimally
nearby paths, according to the variational principle.
Thus the classical limit gives us no guidance as to how to weight
the amplitudes for topologically distinct path sectors relative to one
another.

Let us consider how these remarks
apply to the case of indistinguishable
particles.  In specifying the state of indistinguishable particles,
one can specify only that there are particles at $n$ given positions,
but not any ordering among them.\foot{This should seem obvious on first
hearing, but
startling  upon reflection. After
all, one is accustomed to working with antisymmetric wave-functions
for fermions, and clearly the value of such a wave-function is ambiguous
in the absence of a preferred ordering.
This is a simple case of a general phenomenon that is fundamental
for the physics I am discussing here, to which I shall
return shortly.}
Thus in calculating transition
amplitudes we are interested in
paths that begin at $n$ given points
$(r^{(I)}_j)$ and end at $n$ given points $(r^{(F)}_j)$, which may
or may not be different from the initial ones.
There is one quite obvious way in which
these paths do fall into topologically distinct
sectors.  That is, if one path consists of world lines which
connect the $r^{(I)}_j$ to $r^{(F)}_{{\rm P}(j)}$ with some permutation
$P$, and another path consists of world lines which give a different
permutation, then the paths certainly cannot be continuously deformed
one into the other.

Let us specialize for the moment to just two particles.  Then
of course there are just two distinct classes of permutations.
If we suppose that the only topologically distinct classes of paths
are those that differ by a permutation, we can pin down the
possibilities by using \composing , as follows.  The
classical amplitudes (that is, the amplitudes
formed according to Feynman's prescription using the
classical Lagrangian) automatically satisfy \composing , and
the question is what additional, intrinsically quantum-mechanical,
weights we can apply to the different topological sectors.
We must
obtain the same amplitude by considering the overall composite
process where the two world-lines go through the
sequence $(A,B) \rightarrow (C,D) \rightarrow (E,F)$ as for the
sequence $(A,B) \rightarrow (D,C) \rightarrow (E,F)$, since (by
hypothesis) world-lines of these types
do not differ in {\it overall\/} topology.  But this means that
the extra weighting factor must give unity upon squaring.  Thus
this factor must be $\pm 1$.  These two possibilities correspond, of
course, to the familiar cases of boson and fermion statistics.

One fine point:  I said the
fundamental objects were transition amplitudes between states, but
in fact physics really makes do with
a little less.  We actually only need
that the product of an amplitude
$A( I\rightarrow F; \lambda_1 )$ for one path with
the complex conjugate
$A( I\rightarrow F; \lambda_2 )^{*}$ of the amplitude for another
should be unambiguously defined,
since it is only such interference terms
that ever appear in physical expectation values.  This is important
even for fermions, where
there is a possible ambiguity in defining ``direct'' versus
``exchange'' processes, when
the initial and final configurations do
not consist of the same points.  Fortunately, if we
change our mind as to which is
direct and which exchange we only change an overall phase common
to
both amplitudes, which is not physically significant.

In the foregoing I have implicitly assumed that the
only topological distinction among paths is captured by the
permutation induced by their component world-lines.  This
assumption is correct
(and intuitively obvious)
in three or more space dimensions, but turns
out to be false for two space dimensions, even for two particles.
Indeed, the number of times the world lines wind around one another
as one follows the path from initial to final configurations, is
a topological distinction.  (We also accept
half-windings, for the exchange process.)

Two subtleties deserve to be addressed before this
proposed distinction is accepted.  First, one might be concerned that
the topology is ruined by the possibility that world-lines cross.
This is not true, however, because the crossing itself is a
discrete identifiable event, and one can attach a multiplicative
phase factor (supplying the difference between the two topological
classes) to its occurrence.  Second, the number of windings is not
really well defined if the configuration of initial points is not the
same as the configuration of final points.  This is closely
related to the  problem
we encountered above for fermions, and its resolution is
the same:
the ambiguity affects an overall phase common to all the paths, but
does not affect any physically significant expectation value.
To resolve the
ambiguity {\it does\/} however require us to make a
small adjustment in our
concept of a wave-function.
In the present context, indeed,
the wave-function is not uniquely defined
as a function on the configuration space.
It has a sort of gauge invariance,
in that by changing our convention as to the which braiding (to use the
technical term) is regarded as the reference braiding, we may change the
overall phase of all the amplitudes, but nothing physical.  For
fermions this was only a two-valued ambiguity, and can be absorbed by
the artifice of making the wave-function antisymmetric.  This implicitly
involves a two-fold covering of the identical particle configuration
space. As we shall soon see, things are not always quite so simple.

Acting within the richer topological structure of the path space for
two-space dimensional systems,
the implication of \composing\ is less severe
than in higher dimensions.  Since a path with
$r$ windings followed by a path with $s$ windings yields a path with
$r+s$ windings, it is consistent to add the phase factor $e^{2ir\theta }$.
Bosons correspond to $\theta \equiv 0$ (mod $2\pi$), and fermions
correspond to $\theta \equiv \pi$ (mod $2\pi$).
There is a continuous range of possibilities in between, classified by
the angle $\theta$.  Particles
with these quantum statistics are generically labeled
{\it anyons}.   For anyons, the wave-function must keep track of the
number of windings.  One way to do this is to let the relative angle
$\phi$ between the particles
run from $-\infty \rightarrow \infty$, and impose the condition
$$
\psi (\phi + \pi ) ~=~ e^{i\theta } \psi (\phi )~.
\eqn\wavefunc
$$

\section{Embodiments and Extensions}

It is straightforward to extend these considerations from
$2$ to $n$ particles.  In three or more space dimensions, the
paths fall into classes labeled by the permutation group on
$n$ letters, with composition of paths corresponding to composition
of amplitudes.  \composing\ tells us that in determining the relative
weightings of the different classes we are free to choose a
one-dimensional representation of the permutation group.  There are
exactly two possibilities for this, which of course
just give us the familiar cases
of bosons and fermions.

In two  space dimensions, the topological classification of the paths
is more  complicated.  The group of these classes, with the group
operation corresponding to composition of paths, is the mathematical
object known as the braid group.  The braid group
is a complicated non-abelian group,
but its one-dimensional representations are quite simple.  They are
labeled by an angle $\theta$, which is essentially the angle we already
saw in the 2-particle case.  To get the weighting factor for a given
path one adds the windings for every pair of world-lines to find
the total winding $r$, and multiplies
by $e^{2ir\theta }$.

The wave function for a many anyon system will be subject to
condition \wavefunc\ for every relative angle.  The simple appearance
of this condition is totally deceptive, because the relative angles
form an awkward, partially constrained and partially redundant, set of
variables.

Several generalizations of the original concept of
anyons are possible.  One may allow
the additional weighting factors $A$ to be matrices.
This involves,
in view of \composing , higher-dimensional representations of the
permutation or braid groups.  Higher dimensional
representations of the permutation group lead us to
the old concept of {\it parastatistics}.
There is a serious problem with these kinds
of quantum statistics, however, which perhaps
renders dubious the possibility of their physical realization
in homogeneous many-body systems.  The
difficulty
is, that the higher-dimensional representations do not interpolate in
a
smooth homogeneous way as one changes the number of particles.  If one
has, say, an $n$-particle system, and focuses on different $p<n$-particle
subsystems, one will generically
find that these subsystems are governed by different
rules.

A
more fruitful way to proceed
is to consider a number of different {\it species\/}
of particles, distinguished say by internal quantum numbers $\alpha$.
Then one can have separate numbers $\theta_{\alpha\beta}$ for each
pair of indices, and supply amplitude factors accordingly.  This small
formal step brings up what to me was a surprising point, that in
the circle of ideas around quantum statistics the notion of
indistinguishable particles is not necessarily central.  Indeed the
{\it mutual statistics\/} between different species, parameterized by
$\theta_{\alpha\beta}$ with $\alpha\neq \beta$ is on exactly the same
formal footing as the ordinary quantum statistics parameterized by
$\theta_{\alpha\alpha}$.  This
sort of generalized statistics is quite commonly
realized in the quantum Hall effect complex of states of
matter.

One may also consider, along these lines, the possibility that
the act of
winding one type of particle around another changes internal quantum
numbers.  This situation arises
in simple non-abelian generalizations
of the type of field theories that in their abelian form lead
to anyon and mutual statistics.
The important homogeneity property which fails for traditional
parastatistics will work here.  I believe very rich mathematical
structures arise in this way, but so far this generalization has
not received much attention, and neither the classification nor the
physical analysis of non-abelian statistics has been
sufficiently developed.

So far we have discussed only ideal point particles.  Extended objects
provide important examples of anyons both in
reality and in model field
theories, as we shall see.
They are anyons in
the sense that the relative amplitudes for paths where
the extended objects remain well separated are weighted by phase factors
as if they were point anyons.  In favorable cases
this may be sufficient information
to determine important
aspects of their dynamics at long wavelengths.

Finally I should mention what has been a very active area of
investigation recently, that is quantum statistics in {\it one\/}
space dimension.  There is an element of arbitrariness in defining such
statistics.  Indeed
the topology in path space simply counts the number of leftward
minus rightward world-line
crossings, and a phase at each crossing is just a form
of scattering amplitude, distinguished from the effect of
a local interaction only by its energy independence.  However there are
interesting model systems, describing for instance the
boundary excitations of 2-space dimensional anyon systems,
where one particular choice of statistics is  obviously appropriate, and
leads to an amazingly complete yet fairly
simple description of the dynamics and statistical
mechanics of these models.

\chapter{Dynamical Realization of Statistical Transmutation}

\section{Quantum Mechanical Construction}


So far we have worked at the level of quantum kinematics, and
established at that level the possibility in principle of exotic
possibilities for quantum statistics.
To put some flesh on these bones, we need more substantial dynamical
constructions.

The central physical idea underlying the exotic 2-space statistics is
that particles may acquire intrinsically
quantum mechanical phase factors as they
wind around one another, proportional to the winding.  This is highly
reminiscent of an effect physicists are quite familiar with, the
famous Aharonov-Bohm effect.  Let me remind you that this is the effect
that although outside an infinitely thin solenoid there is no magnetic
field, and therefore no classical force, there is still
a vector potential, and this vector potential can dramatically
affect the behavior of charged particles in quantum mechanics.  In fact
the vector potential affects the particles by imparting to their
wave-functions a trajectory-dependent phase
$$
\exp \bigl( iq \int^{r^{(F)}}_{r^{(I)}} a\cdot v dt \bigr) ~=~
\exp \bigl( iq \int^{r^{(F)}}_{r^{(I)}} a\cdot dr   \bigr)          ~.
\eqn\abphase
$$
One will have non-trivial interference effects due to these phase
factors,
if they give different phases for different trajectories with the
same end-points.  In the case at hand, the relative value
of \abphase\ for two trajectories
$T_1,T_2$
is simply the exponential line integral
of the vector potential over the closed path $(T_2)^{-1}T_1$
obtained by following
$T_1$ from $r^{(I)}$ to $r^{(F)}$  and the reverse of $T_2$ from
$r^{(F)}$ to $r^{(I)}$, and this is
$$
{\rm relative~ phase} ~=~ \exp (iq \oint a\cdot dr) ~=~ \exp (iq \Phi w)
\eqn\relphase
$$
where $w$ is the winding number of this path around the solenoid and
$\Phi$ is the magnetic flux the solenoid encloses.

\relphase\ is exactly the sort of factor we need to implement the
fractional statistics.  Thus we can transmute
the statistics of two-space
dimensional particles by imparting to them fictitious charge and flux.
 From this vantage point, the crucial distinction between two and higher
space dimensions is that in two dimensions flux {\it tubes\/}
become flux {\it points}, so one has to take into account
possible interactions due to phase factors of the Aharonov-Bohm type
even in the dynamics of point particles.

\section{Field Theoretic Realizations}


 From these elementary considerations it is only a short step to
what I believe is the most profound and useful way of viewing the
whole circle of ideas around anyons, that is to see them as realizations
of a special kind of gauge symmetry.  Indeed, we can construct a field
theory that implements the attachment of fictitious charge and flux
to particles, as follows.  Add to the standard point-particle action
in two space dimensions the interaction terms
$$
\Delta {\cal S} ~=~
  \int d^2xdt (qj\cdot a +
    \mu \epsilon^{\alpha\beta\gamma} a_\alpha \partial_\beta a_\gamma )~.
\eqn\modL
$$
In this expression,
$j^\alpha(x,t)~=~ \Sigma_k {dx_k^\alpha \over dt} \delta^2(x - x_k(t))$
is the
particle number current.  The last term, multiplying $\mu$,
is called
the Chern-Simons term after the mathematicians who discussed related
concepts
in quite a different, differential-geometric, context.  Its existence,
in this specific form, is obviously closely tied to the 2+1 dimensional
nature of the models we are considering.

$a$ is of course to be distinguished from the ordinary electromagnetic
field, for which we shall reserve the symbol $A$.  $a$ is supposed
to occur nowhere else in the action but only in the terms displayed in
\modL ; in particular it has no Maxwell kinetic term.
The equations of motion obtained by varying the $a_\alpha$ are
$$
q j^\alpha ~=~ \mu \epsilon^{\alpha\beta\gamma} f_{\beta\gamma}~,
\eqn\eqom
$$
where $f_{\beta\gamma}~=\partial_\beta a_\gamma - \partial_\gamma a_\beta$
is the field strength.  These are evidently invariant under
the gauge transformation
$a_\alpha \rightarrow a_\alpha + \partial_\alpha \lambda$, as we might
have anticipated from \modL . The $\alpha = 0$ component of the
equation connect the particle density $\rho$ to the fictitious magnetic
field $b$ according to the fundamental relation
$$
q \rho ~=~ \mu b~.
\eqn\basiceq
$$
This is exactly the flux-attachment prescription
we discussed before:
the charged particles have become flux points as well.
The $qj\cdot a$  terms
insure that the particles do indeed interact as
ordinary charged particles with respect
to the fictitious field $a$, and so \modL\ has indeed captured the
physics of statistical transmutation.  Adding \modL\ to a given
point-particle Lagrangian, with a conserved current $j$, will
change the quantum statistics by
$$
\Delta \theta ~=~ q^2/\mu~.
\eqn\delthet
$$

It is instructive briefly to consider how a Hamiltonian treatment
would go.  Since $a_0$ does not occur with time derivatives,
\basiceq\ emerges as a constraint.
Integrating by parts, we can also arrange that $a_1$ does
not appear with time derivatives.  Then its equation of
motion generates a constraint
that allows one to express $a_2$ in terms of the particle variables.
Thus in principle by introducing \modL\ we have not added new degrees
of freedom, but only modified the properties of the existing ones.
Explicit elimination of the $a_\alpha$ in terms of the particle variables
leads to unwieldy non-local expressions, however, so that if the
fictitious field variables behave in a particularly simple or transparent
way there may be
great advantage to introducing them explicitly and keeping
them in the foreground.  As we will soon discuss, this is indeed the
case.

Before entering into the applications, however, let me briefly mention
how the various generalizations we previously contemplated at the level
of kinematics can be realized in local gauge field theories.  It is
quite clear how to introduce mutual statistics.  If there are several
conserved particle currents $j_{(\mu)}$, and several
gauge fields $a_{(\sigma)}$ one can transmute both self-
and mutual statistics by adding to the action terms of the type
$$
\Delta {\cal S} ~=~
  \int d^2xdt
  \bigl( \Gamma_{(\lambda) (\mu) (\sigma)} q_\lambda j_{(\mu)} a_{(\sigma)}
  + N_{(\sigma)(\tau)}
  \epsilon^{\alpha\beta\gamma} a_{(\sigma)\alpha}
  \partial_\beta a_{(\tau)\gamma} \bigr)~,
\eqn\genaction
$$
$\Gamma$ and $N$ are numerical arrays, and we may suppose
that $N$ is symmetric.  Such generality
may seem like overkill, but in fact fairly
elaborate theories of this kind are used
to describe observed fractional
quantized Hall effect states.

What about extended objects --
including field, as opposed to point-particle,
theories?  In this direction
too, once the simplest case is before us it is clear how
to generalize.  To make the statistical transmutation construction, we needed
a conserved current.  Indeed, only when we have a conserved charged will
be able to identify in an unambiguous way ``what it is'' that is having
its quantum statistics transmuted.

There are actually two cases to consider.  If the
current in question has a topological character, and is conserved independent
of the equations of motion, then we can just take over \modL\
for it directly.
A nice example is the nonlinear $\sigma$-model of a field
$\vec n$ with three internal components subject to the constraint
$(\vec n)^2 =1$.  It has the topological current
$$
j^\alpha ~=~ \epsilon^{\alpha\beta\gamma} \epsilon_{abc}
            n^a \partial_\beta n^b \partial_\gamma n^c~.
\eqn\topcurr
$$
The statistical transmutation construction will alter the statistics
of the solitons which carry this conserved charge, that is the
so-called ``baby Skyrmions''.  In this case, if one eliminates the
fictitious gauge field one finds that what one has added
to the Lagrangian is a non-local integral representation of the
Hopf invariant, a classic of topology.

More common is the case where the conservation law is associated with
an internal symmetry.  In that case, we can
just follow the minimal coupling
prescription to replace the $j\cdot a$ with a proper gauge-invariant
generalization, and proceed as before.  The quanta carrying the quantum
numbers of the conserved current will have their statistics transmuted.
The fictitious gauge fields will again not represent independent degrees
of freedom, but rather a way of encoding properties of the particles --
or rather, in this generality, the fields.

Non-abelian generalizations of these constructions are also possible.
Many subtleties arise when one considers their global
properties, but at the level of local equations of motion there is
just one, and even it is quite simple.  That is, the Chern-Simons
term goes over into
$$
\Delta {\cal S}_{\rm CS} ~\propto ~
  \int d^2x dt  \epsilon^{\alpha\beta\gamma}
    {\rm Tr} (a_\alpha \partial_\beta a_\gamma
      + {2\over 3} a_\alpha [a_\beta, a_\gamma] )~.
\eqn\nacs
$$
In this expression a matrix representation of the fields has
been assumed, and Tr indicates the trace.
The funny ${2\over 3}$ is necessary so that when one varies to get
the equation of motion the proper gauge invariant field strength,
{\it i.e}.
$\partial_\alpha a_\beta - \partial_\beta a_\alpha + [a_\alpha , a_\beta ]$,
is obtained.

One might also inquire whether exotic statistics might arise more
organically, within
field theories that do not make use of these particular
constructions.  The answer to this question, of course, will also
color our expectations regarding embodiments of exotic statistics
in reality.

First let me make two preliminary remarks, that perhaps I should have made
earlier.  The requirement of two-space dimensionality might seem to
be very restrictive, since real-world systems are at some level
(it would seem) three-space dimensional.  This restriction is easily
breached, however.  The point is that one may consider systems that
are homogeneous in two directions but inhomogeneous in a third.
At sufficiently low energies, it may be impossible to excite motion in
the third direction.  In that case, a two-dimensional description is
strictly valid.  Of course much of modern electronics takes place in
an essentially two-dimensional world, and many of the most interesting
interactions between material bodies takes place at their interfaces, so
essentially two-space dimensional phenomena can be of direct physical
interest.

Second, except in the cases
of bosons and fermions quantum statistics appears to violate
P and T symmetry, since the phase associated with world-lines
winding one way is not the same as the phase associated with world-lines
winding the opposite (that is, P- or T-reversed) way.  Thus it would
seem that to have exotic statistics we need to have these
discrete space-time symmetries
broken.
Such symmetry breaking
is of course not rare in condensed matter systems:
it is the essence of magnetic order.
Actually, though, it is not even strictly necessary.   Even within an
overall P- and T- conserving theory, there may be particles which
are not P and T eigenstates.  Thus for example
a magnetic flux-point
in two space dimensions, like a magnetic monopole in three space dimensions,
is {\it not\/} a state of definite P and T -- rather, those symmetries
take these particles into their antiparticles!  It might seem paradoxical
that one should work with  eigenstates of a P- and T- conserving Hamiltonian
which are not eigenstates of P and T, but it is natural here because these
operations do not commute with a quantity -- that is, in both cases, total
magnetic flux -- which defines a superselection sector.
Thus if we work -- as experimentalists are forced to --
within a definite superselection sector, the
symmetry is essentially useless.  Similarly
although naive P and T
transformations change the sign of the Chern-Simons term in \genaction\
({\it i.e}. of the matrix $N$) in modified form
they may still define valid symmetries, after being
combined with a suitable transformation in internal space.
Thus in general one may expect that exotic statistics
is a valid possibility
for the effective low-energy theory of excitations even within
P- and T-conserving theories, in theories that have non-trivial
superselection sectors.

 From the point of view of general principles of effective theory, a
crucial aspect of the terms in \genaction\ is that they are of very
low dimension.  They involve only a small number of fields,
and very few derivatives.  Thus they are ``relevant'' operators in the
technical sense, and if there is no symmetry to preclude them we
must expect that they will control the physics at long wavelengths and
low energy.  Experience with integrating out the high-energy degrees of
freedom in model  2+1 dimensional field
theories bears this out, in that a Chern-Simons
interaction often appears in the resulting low-energy effective field
theory.  Similarly, a dynamical calculation for
quasiparticles in the fractional quantized Hall effect shows that they
obey anyon statistics.
And for
flux tubes in non-abelian gauge theories
it requires a conspiracy
to {\it avoid\/}
expressing exotic
statistics.

\chapter{The Master Phase Diagram}

\section{The Flux Trading Paradigm}

A simple but powerful principle allows one to put the
abstract idea of
statistical transmutation to work as a tool for understanding the
properties of two space-dimensional many-body problems.
It was first mentioned as a possibility by Arovas, Schrieffer,
Zee and myself in the same paper
where the gauge theory of fractional statistics was
first clearly formulated, but its real power only became clear
when Laughlin used it derive anyon superconductivity.  A related,
but (to my mind) less rigorously formulated idea helped motivate
Jain to his remarkable constructions of trial wave functions for
the fractional quantum Hall effect.  Greiter and I tried to put
the essential idea in as clear and simple a form as possible, and it
is that version I will show you today.  I will also be alluding to
very important subsequent work of Halperin, Lee, and Read, whose theory
of compressible Hall states is probably the most impressive work in
the field so far.

The principle is simply this: that it may be a valid approximation,
in calculating certain low-energy properties,
to replace the flux attached to a collection of anyons by its uniform
average.  Two questions will no doubt occur to you: why is this
at all reasonable? and what is it good for?  Let me answer these in
turn.

What we are proposing is that it can be valid to replace a bumpy
distribution of small amounts of flux attached to particle positions
with a smoothed-out uniform flux.
Intuitively, we might expect that
quantum-mechanical uncertainty in position will tend to
smooth out the particle positions -- in particular, one cannot
localize particles in a strong magnetic field better than to within the
magnetic length $l_B = \sqrt {hc\over qB}$).  This argument
cuts twice,
in that we care about the effect of this field on the particles themselves,
so that there is smoothing both from the source and from the probe.

A more correct and powerful argument, however, goes along quite different
lines.  Let us suppose that the change in statistics is small.
Why not just treat it as a perturbation, full stop, without bothering
about the uniform field?  The difficulty with this procedure
is that even a very small change in statistics produces a numerically
large perturbation, for the following reason.  A change in statistics
is associated with a gauge field, as we have seen, whose vector potential
must reflect the fundamental constraint \basiceq .  By integrating
this constraint around a large circle, we see that the azimuthal
vector potential
must behave as
$$
a_\phi (r)~\sim~ {q\over 2\mu} \rho r~,
\eqn\vecpot
$$
using unit normalized (Cartesian) basis vectors.
This vector potential grows as the size of the system, and clearly
for a large system it cannot be treated as a small perturbation.
However,
we may expand the Hamiltonian as
$$
H ~=~ H_0 + (H-H_0)~,
\eqn\hamexp
$$
where in $H_0$ we use a uniform magnetic field of strength
$q\rho/\mu$ instead of the true $a_\phi$, and in $H-H_0$ restore the
residual terms.  In this way we may hope to
arrive at
a tractable problem for $H_0$ -- which is still fairly simple,
since it involves particles moving in a uniform background field --
and
a perturbation to it that is genuinely small, because the
dangerous
long-range pieces, which would show up as infrared
divergences, have been subtracted off.

One expects that generically a small regular perturbation of a many-body
system with a gap in its energy spectrum
will not destroy the gap.  Thus one expects, from the argument I just
outlined,
that if one can trade statistics for uniform magnetic field along
the lines
$$
\Delta {1\over \nu } ~=~ \Delta {\theta \over \pi}
\eqn\master
$$
in the magnetic field-statistic plane, where
$$
\nu ~\equiv~ {\rho \over qB }~,
\eqn\fillfrac
$$
the so-called filling fraction,
is the ratio of the actual density to that necessary to fill one
Landau level.   More precisely, one expects that generically as one
moves along such lines states with energy gaps -- incompressible states
-- will be related to other, qualitatively similar incompressible states.

In writing \fillfrac , I have snuck in a crucial physical point
by notational sleight-of-hand.  Whereas the preceding discussion
has related to fictitious fields, the uniform field we finally used in
our approximation has exactly the same effect as an honest-to-goodness
applied magnetic field.  Thus we are approximating anyons of one type
in a given uniform magnetic field by anyons of another type in a different
uniform magnetic field, as is implicit in \fillfrac .

\section{A Tour of the Quantum Hall Complex}

With this background, I can now present the centerpiece of this
talk, Figure 1, which I call the Master Diagram.  It shows in what
I hope is clear and easily comprehensible form how the
simple ideas just outlined can be used to give a coherent account of
several different exotic states of two-dimensional matter
-- a ``grand unified theory'' of the quantum Hall complex.
I would like now to take you on a brief tour of the Master Diagram.

If we start from fermions and gradually but monotonically
change the statistics, we
will eventually arrive back at fermions.  Of course, the final
state in this process, being a state of fermions,
is a candidate state
to describe electrons again.  In the new state the
electrons will be in a different
magnetic field, and may have acquired
very different correlations in the course of the
adiabatic flux-trading procedure,
but if the initial state has an
energy gap
the final state has a good chance of sharing that property.

Let us first apply this procedure
starting with a state for which the existence
of an energy gap can be argued on very general grounds, that is a single
filled Landau level ($\nu=1$).  Moving along the
$\Delta (1/\nu )= \Delta (\theta/\pi )$, we reach fermions again when
$\Delta \theta = 2\pi$, that is $\Delta (1/\nu) = 2$ or
$\nu = 1/3$.  One may continue along this line to reach
$\nu = 1/5,1/7 ...$
-- the filling fractions described by the classic Laughlin
theory of the fractional quantized Hall effect.
The incompressibility of these states is a true many-body property,
that
is just plain
false in the absence of interactions.  Indeed, there is a vast
degeneracy for non-interacting fermions  partially filling a Landau band,
reflecting the many choices available for
which single-particle states to fill.  It is quite remarkable that
one is led to anticipate the incompressibility of these states from
such simple arguments, based only on the stability of what appears
to be a controlled perturbation theory.
Although I am not the person to do it, I feel confident that
a rigorous
proof of the incompressibility of
simple fractional Hall effect states, which
has so far been elusive, could be constructed along these lines.

If we start with two filled Landau levels and follow the same logic
we arrive at filling fractions $2/5, 2/9, ... $; and so forth.
In this way, a large number of fractional quantized Hall states can
be built up.
This is closely related to most elementary version of constructions
carried to a high level of sophistication by Jain (and backed up by
some brilliant analytic and numerical work).
The standard hierarchy construction can also be discussed in this framework,
if one combines the flux-trading construction for elementary particles
with a ``dual'' construction applying transmuting
vortices of the gauge fields
on  preceding levels and iterates.
These vortices of course have a conserved
topological current, and are therefore subject to the
transmutation construction as previously discussed.

Quite a different use of the same idea is to extrapolate along these
lines toward
zero magnetic field, rather than toward Fermi statistics.  If one
hits zero magnetic field while maintaining a gap in the single particle
spectrum, one has produced a candidate superfluid.  For at zero magnetic
field there is no longer a preferred density, and one expects a collective
mode of density waves whose energy should go to zero at long wavelengths.
These are
stable according the Landau criterion -- because there is
an energy gap for ordinary quasiparticle excitations, decay
of the collective mode is forbidden by energy-momentum
conservation.  According to
modern understanding one must establish
somewhat more than this for true superfluidity, and there is
some controversy whether
one has true superfluidity for these states,
though I at least am convinced one does (to the extent one can
in two space dimensions; that is, in the sense of Kosterlitz-Thouless
theory).
In any case something very interesting happens, and we can identify the
statistics for whatever it is occurs.
Starting from $n$ filled Landau levels,
one extrapolates according to the master formula \master\ to zero magnetic
field for statistics $\theta/\pi = 1-1/n$.  Ideal quantum fluids of
particles with these statistics are predicted to be superfluids at zero
temperature, a concept which has received much support from expansions
in $1/n$, random-phase and Hartree-Fock calculations, and numerical work.
A particularly interesting case is $n=1$, which corresponds to
bosons!   One does indeed expect the hard-core Bose gas to form
a superfluid, but it is perhaps
somewhat startling to trace the ancestry of this
classic superfluid to a
filled Landau level.   The case $n=2$, or semions, arises in some ideas
about high-temperature superconductivity, but unfortunately
these ideas have failed the test of experiment, at least in their simplest
form.

Finally, following Halperin, Lee, and Read, let us consider extrapolating
from fermions in {\it zero\/} magnetic field, that is $\nu = \infty$.
This is a bold thing to try, because the starting point is not a system
with a gap, and one is doing degenerate perturbation theory from the
get-go.  However, one has become
accustomed, from Landau's Fermi liquid theory, to the idea that Fermi
surface
is quite robust against (repulsive) perturbations.
Roughly speaking, this is because although there is no gap there is
effectively a very small phase space for interactions, due to the
exclusion principle.  The vanishing phase space over-cancels the
vanishing energy denominators.
Extrapolating, then, according to \master\ one
arrives back at fermions for $\nu = 1/2$.   One is led, therefore, to
predict that there should be Fermi surface effects at $\nu = 1/2$,
notably including the existence of
a wealth of low-energy fermionic quasiparticles that propagate in
straight lines (the eigenstates of $H_0$, in the sense of \hamexp ).
Remarkably, several such effects have indeed been observed, and agree
at
least semi-quantitatively with the theory.

In view of the boldness of the extrapolation in this last application,
one must
examine the effect of the restoring the dynamics of the fictitious
gauge field, {\it i. e}. of applying $H - H_0$ to the putative ground
state.  An adiabatic approach is impractical, since the Pauli exclusion
principle is not built in at intermediate stages, so to do any explicit
calculation it appears one
must resort to perturbation theory.
Since the perturbation is in no sense small numerically  --
the statistics
has been changed by $2\pi$ -- special techniques must be used.  Nayak
and I,
using an unusual renormalization group argument,
found that at low energies (as the
quasi-particle energies approach the Fermi surface) one is
led toward small effective couplings, but only logarithmically.  This
justifies the free theory as a self-consistent
approximation at low energies, but also
indicates its fundamental limitations.
Some further discussion of these
ideas, including a proposal for a crucial experimental test, is given
below.

\chapter{Properties of the Quasiparticles}

Some of the most interesting properties of the states in the
quantum Hall complex relates to the elementary excitations about them.
I shall now describe how some of these properties are manifested
directly in elementary space-time processes [\disclaim ] .

\section{Existence and Peculiarities}

The localized particle-like character,
and the fractional charge and quantum statistics, of the
quasiparticles in the fractional quantized Hall effect is
a very plausible consequence of the
concepts we have
discussed in connection with the Master Diagram.
The localized nature of the allowed density fluctuations follows
from the incompressible nature of the state: to create an
inhomogeneous charge one needs to exceed a definite threshold, and
above this threshold
one creates quasiparticles in a localized region.  This is
the only option, because
the bulk state has a preferred density, and it would cost a finite energy
per unit area to vary from that density.
Moreover
in moving from the single filled Landau level to the $\nu=1/3$ state,
for instance, we cranked up two extra units of quantized
flux on the electrons.  These, when added to the effective single
unit of flux connected with Fermi statistics, yield the result
that a phase
of $6\pi$ is accumulated as one electron circles another.  This provides
a strong hint
that the electron is not the most elementary excitation.
The most elementary excitation, that provides the minimal flux consistent
with not disturbing the boundary conditions on electrons arbitrarily far
away, should be associated with phase accumulation $2\pi$.  Since it
takes three of these to reproduce the effect of an electron, one should
expect that each such elementary excitation carries one-third the
quantum numbers of the electron.   Applied to electric
charge, this argument clearly leads to the assignment
$e/3$ for the quasiparticles.  For statistics the division is not
unambiguous (because the statistical parameter is defined only
modulo $2\pi$) but one might expect that the most economical value,
$\theta = \pi/3$, is realized.  This value is indeed what the microscopic
theory indicates.

\section{Quantum Drifts}

The first chapter of almost
any text on plasma physics contains the
theory of
particle drifts, that is the motion of charged particles
subject to a strong uniform magnetic field
and additional weak fields.  The leading motion is
fast cyclotron motion in circles of radius
$r_{\rm cyc.} = {pc\over qB} $ at frequency
$\omega_{\rm cyc.} = {qB \over mc}$.  The ``guiding center'' of
this circle undergoes slow drifts if there are additional smooth,
weak fields, due to non-cancelation of impulses around the
cyclotron orbit.

In quantum mechanics another parameter of length becomes
available, that is the quantum magnetic length scale
$l_{\rm quantum} = \sqrt {\hbar c\over qB }$.
For very strong fields evidently $l_{\rm quantum} > r_{\rm cyc.}$
and the classical cyclotron orbits lose their meaning.
The drifts must be calculated in a different way, and the results
are different.  For example, since the classical equations of
motion depend only on the charge-to-mass ratio $q/m$ observations
of the classical drifts of a particle
do not allow one to infer a definite value of the charge; but
as it turns out the quantum drifts do.

The quantum drifts are of considerable intrinsic interest, and
of specific interest for
the quantum Hall effect, since the quasiparticles
are predicted to be fractionally charged, and perforce must be
studied in the presence of a large background magnetic field.  They
are well within the quantum drift regime.

Another way of thinking about the quantum drifts is that they
represent the dynamical response of particles confined, in energy
space, to the lowest Landau level.  From this point of view we see
right away that
straightforward perturbation theory is insufficient for this problem,
because the problem is highly degenerate.  One can make progress by
assuming cylindrical symmetry for the perturbing fields.  Then the
perturbing Hamiltonian is diagonal in canonical angular momentum, and
since the lowest Landau level contains just one level for each
non-negative canonical angular momentum one can calculate each
level separately, then sew together wave packets and follow the
dynamics.

I refer you to the reference mentioned in [\disclaim ]
for a full discussion,
including some numerical simulation and computer-generated pictures
of representative motions.

\section{Mass Renormalization and Time-of-Flight}

As mentioned above, Halperin, Lee and Read proposed a striking
description of the compressible Hall state near filling fraction
$\nu = 1/2$, that has gained considerable experimental support.
Their theory is based on the idea of approximating the electrons
by fermions in {\it zero\/} magnetic field.  It is
proposed that correlations among the electrons are such as to
cancel the effect of the imposed magnetic field, so that the
dressed quasiparticles have the quantum numbers of electrons but
travel (to a first approximation) in straight lines.  The special
feature of $\nu = 1/2$ is that attaching notional two-unit flux
tubes to the electrons formally does not alter their properties,
yet it
involves a vector potential
which cancels off the growing part of the real
imposed
vector potential.  One hopes to treat the difference as a perturbation.

The perturbative treatment of the statistical gauge field is intrinsically
limited, however, because the coupling is neither small nor truly
short-ranged.  Nor, since it is essentially a magnetic coupling, is it
effectively screened.  Nevertheless one may hope to justify the
simple theory, similarly to the way one justifies the use of perturbation
theory in QCD, by showing that the {\it effective\/}
coupling can become weak in an appropriate limit.

We are concerned with the long-wavelength
behavior, specifically with the question whether infrared
divergences associated with the deviation of the statistical gauge
field from a uniform constant -- in other words, with its fluctuations --
are suppressed.  This formulation of the
problem suggests the used of a
renormalization group approach.

Now the longitudinal
electric fluctuations of the gauge field should be screened
more or less
in the usual way, and the dangerous fluctuations are the magnetic
ones.  Since in the Chern-Simons theory the magnetic field is
enslaved to density fluctuations, it is important to consider the
effect of additional
long-range repulsive interactions, which tend to suppress
such fluctuations.  For the sake of generality, one considers a
$1/|k|^x$ interaction, where $x =1$ is the realistic
Coulomb case.

Among the coupling in a non-relativistic Lagrangian is the mass
term ${1\over 2m^*} (\nabla \psi )^2$, and it turns out that
the crucial renormalization
for the gauge theory is the renormalization of $m^*$ as one scales toward
low
energies, and momenta on the Fermi surface.  Indeed since the fundamental
coupling is magnetic, an increase in $m^*$ tends to suppress it.
We find a fixed point for the effective coupling at
$\alpha^* ~=~ (1-x)/4$; for the Coulomb case there is a logarithmic
approach to zero coupling.  For this case the effective mass behaves as
$$
m^*(\omega) \rightarrow {\rm const.~} \ln {1\over |\omega |}
\eqn\effmass
$$
as one approaches the Fermi surface at $\omega = 0$.   Notice that this
mass diverges at the Fermi surface, so that strictly speaking the Fermi
liquid theory does not apply, although many of its qualitative features
survive with only small corrections.  Clearly the predicted behavior
\effmass\ is of fundamental interest, as is the
question whether it can be
seen directly in space-time.

Actually to do this would seem to require only that some beautiful
measurements of Goldman, Su, and Jain be extended.  These authors
tested some fundamental properties of the quasiparticles by working
in magnetic fields $B_{1/2} + \delta B$ slightly different from those
which give exactly $\nu = 1/2$.  The quasiparticles are then supposed
to see the effective field $\delta B$, which causes them to move in
large cyclotron orbits.  In fact since the cyclotron radius is
$r=pc/(e\delta B)$ (here the small-field, classical theory applies!)
and the momentum is cut off at the Fermi surface there is predicted to
be a minimum radius, a prediction which was verified experimentally using
a two-slit arrangement.  Now what is required to test \effmass\
is clearly that the {\it time of flight\/} should be measured as a
function of the radius; from it one readily obtains the velocity and then,
knowing the momentum, the effective mass.

\refout

\endpage

\FIG\master{The Master Diagram for the quantum Hall complex, discussed
at length in the text.  It demonstrates that one can
understand many interesting
phases of quantum matter in two space dimensions
in a simple, unified way using the concepts of statistical transmutation
and flux trading.}
\endpage


\epsfysize=.5\vsize\hskip1cm
\vbox to .6\vsize{\epsffile{btalk.eps}}\nextline\hskip-1cm
\endpage

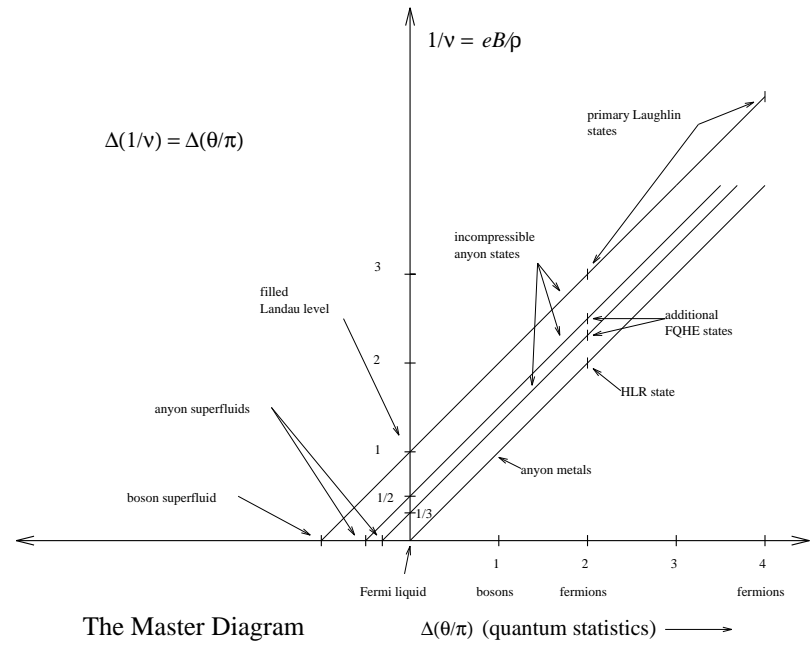
\end